# Intelligent Imaging of Thermal Transport in Nanostructures

Xin Qian[1*], Te-Huan Liu[1], Xiaobo Li[1*] and Ronggui Yang[2*]

[1]School of Energy and Power Engineering, Huazhong University of Science and Technology, Wuhan, China 430074
[2]College of Mechanics and Engineering Science, Peking University, Beijing, 100871, China

Corresponding emails: xinqian21@hust.edu.cn; xbli35@hust.edu.cn; ronggui@pku.edu.cn

## Abstract

The ability to measure thermal properties and visualize heat transport processes with high spatial resolution is becoming increasingly important for materials discovery, next-generation electronics, and advanced manufacturing. As critical dimensions and architectures of novel materials and high performance electronics shrink and become more heterogeneous, thermal transport is no longer governed by bulk material properties alone, but by local variations in microstructure, such as interfaces and defects, while the operating conditions can also play a role. With the development of non-contact pump-probe thermoreflectance techniques, specific heat, thermal conductivity, and interface conductance can be mapped either through wide-field thermos-reflectance images or by scanning technique. However, key challenges remain such as the balance between imaging throughput and measurement accuracy, the extraction of multiple parameters with high confidence, and non-intrusive subsurface characterization, etc. This Review highlights recent advancements in artificial intelligence (AI)-assistance in high-throughput data processing, visualizing interfacial thermal transport, and nonintrusive subsurface analysis, and offers outlooks for future research directions such as spectral mapping of heat carriers, three-dimensional property mapping, and AI-agentic experimental designs. We anticipate this Review will motivate future efforts in combining AI with experimental techniques to achieve unprecedented capabilities for thermal characterization of nanoscale materials and device characterization, facilitating materials discovery and electronics revolution.

## 1. Introduction

Resolving the thermal transport properties with nanoscale resolution is becoming increasingly pivotal in modern technologies. In electronics, the continuing scaling of transistors and computational power has led to unprecedented heat generation[1-2], resulting in diminished benefits of continuing the Moore's law. The pursuit of higher system-level performance and the AI momentum have been driving the heterogeneous and three-dimensional integration of functional electronics [3], resulting in many more interfaces and increasingly nonuniform heat generation. Similar challenges arise in miniaturized energy-conversion systems, including batteries[4], thermoelectrics[5-6], photovoltaics[7], where local variations in microstructure, interfaces, defects, and device architecture[8] strongly affect the thermal transport processes.

Understanding how heat flows in micro/nanostructured materials and devices requires the ability to measure and visualize heat transport with high spatial resolution, which enables identification of thermal bottlenecks within complex materials and operating devices [9-13]. Over the past three decades, ultrafast laser-based pump-probe thermoreflectance methods, primarily the time-domain thermoreflectance (TDTR)[14-15] and the frequency-domain thermoreflectance (FDTR)[16-17], have evolved as powerful tools for characterizing thermal properties of thin films [18-20] and bulk materials, including wide-bandgap semiconductors[21-23], van der Waals layered materials[24-26], superlattices[27-28], as well as heterogeneous interfaces [18, 29-30]. Compared with conventional thermal metrology, pump-probe thermoreflectance techniques offer microscale spatial resolution and high time resolution (ps to μs), enabling determination of multiple thermal properties including specific heat [31], anisotropic thermal conductivity[32-34], and interfacial thermal conductance[35] by tailoring the beam shape, spot sizes, and modulation frequencies of the pump-lasers.

Motivated by the pivotal role of visualizing thermal transport properties in uncovering structure-property relations and providing guidelines for design and optimization of materials, interfaces, and devices. Significant efforts are pursued in pushing forward thermoreflectance methods for imaging thermal transport with high resolution and high throughput [36-38]. For example, scanned images of thermal conductivity across the sample surface directly leads to the identification of ultrahigh intrinsic thermal conductivity (> 1000 $Wm^{-1}K^{-1}$) of low-defect-regions in cubic boron arsenide[39-41], isotope-enriched cubic boron nitride[42], and θ-tantalum nitride [43]. Phonon interaction with grain boundaries can be visualized by mapping the local thermal conductivity in polycrystals[44-45], manifested as suppression of local thermal conductivity near the grain interfaces. Property profiling over gradient samples provides an efficient and cost-effective way of investigating size or composition dependence for material development. For example, the size-dependence in thermal transport can be effectively measured by scanning across a wedge-shaped thin film rather than repetitively measuring samples with varied thicknesses[19]. For alloy systems, thermal conductivity across the entire phase diagram can be characterized by imaging thermal properties over a wedge-shaped combinatorial sample with gradient chemical compositions[46-48]. On the other hand, imaging thermal conductance of buried interfaces can help examining the bonding quality for wafer-scale integration[49-50]. These thermal property images, if archived and organized properly, can serve as high-fidelity datasets for training artificial intelligence (AI) models for material discovery and device design [51-53].

However, there are still many key challenges in pump-probe thermal property microscopy. First, it is difficult to simultaneously achieve high imaging speed, measurement accuracy, and resolution. To achieve high-resolution, the scanning interval must be set as small as possible, resulting in a large amount of data to be processed. The imaging speed can be enhanced by measuring less datapoints at each location, which in turn results in higher uncertainty. Second, pump-probe thermal property imaging involves complicated inversion problems in data processing. Multiple fitting parameters can couple with each other depending on the experimental parameters, such as delay time, modulation frequency, and laser spot sizes [54-55]. In addition, when constructing high-resolution property maps, adjacent measurements can "cross-talk" with each other. Due to the finite sizes of pump and probe beams, each imaging pixel measures an averaged property within the beam area. If the distance between neighboring measurement locations becomes smaller than the laser spot size, there could be crosstalk due to the overlapped sensing area, which is usually neglected in conventional pump-probe thermal models for single-point measurements. Third, profiling thermal property in the depth direction with good resolution is challenging when scanning across the sample surface. In pump-probe experiments, temperature rise is detected at the sample surface, which is an integrated result of heat dissipation at all locations within the sensing depth of a few times the thermal penetration length [56]. Usually a constant thermal conductivity is assumed to fit the experimental signal, which apparently fails if the material structure or composition varies in the depth direction, such as chemical-vapor-deposited polycrystalline diamond[57-58], ion-irradiated semiconductors[59], and thermal management materials[60]. There are efforts in modulating sensing depths to extract depth-dependent thermal properties; however, one still needs to assume a function of the depth dependence with fitting parameters [59], which requires prior knowledge about how thermal properties are distributed. However, information about the depth profile is usually not available before performing measurements.

With the increasing momentum in AI for science [61-62], the past five years have seen preliminary efforts in integrating machine learning (ML) with pump-probe thermal metrology to address the aforementioned challenges, leading to enhanced capabilities in thermal property mapping. This Review aims to highlight the importance of combining AI and pump-probe experiments with the unprecedented capability for thermal property visualization. This Review is organized as follows. First a brief introduction to the fundamentals of the pump-probe thermoreflectance imaging technique is discussed, and we then highlight the recent advances in enhancing imaging capabilities of thermal transport in nanostructures by integrating the pump-probe experiments with AI/ML algorithms. Finally, we provide outlooks for a few promising future directions for combining AI with thermal characterization to achieve unprecedented measurement capabilities.

## 2. Fundamentals of pump-probe thermoreflectance microscopy

Pump-probe thermoreflectance is a powerful optical characterization technique used to measure the thermal properties of both thin film and bulk materials, which are typically coated with metallic transducer films. The fundamental principle relies on using a modulated pump beam to heat the sample surface, creating a transient temperature change $\Delta T$, while the probe beam measures the corresponding change in the surface's reflectance. Since the reflectance of metal transducers coated on the samples changes with $\Delta T$, the temperature evolution dynamics is embedded in the probe signal. The pump-probe thermoreflectance technique is most often implemented in two ways. TDTR splits an ultrafast pulsed laser into separate pump and probe paths, with a mechanical delay stage to control the probe’s arrival time relative to the pump pulses, and the TDTR signal is detected as a function of the delay time (Fig. 1a). Instead of controlling the probe delay time, FDTR does not require a pulsed laser source and detects the thermoreflectance as a function of the modulation frequency (Fig. 1b). More detailed discussion on working principles and experimental implementations of TDTR or FDTR can be found in previous tutorials or reviews [2, 15, 63]. In both methods, thermal properties are extracted by fitting the experimental signal using a thermal diffusion model assuming an equilibrium local thermal conductivity. The thermoreflectance response in the frequency domain is calculated as [14, 17-18]:

$$H(\omega) = \beta \int_0^{\infty} \mathcal{G}(\omega, \xi) \cdot \exp\left(-\frac{w_{rms}^2 \xi^2}{4}\right) \xi d\xi \tag{1}$$

where $\beta$ is a proportionality factor determined by the thermoreflectance coefficient, gain of detectors, and laser power; $w_{rms} = \sqrt{(w_0^2 + w_1^2)/2}$ is the root-mean-squared $1/e^2$ beam radius of the pump ($w_0$) and probe ($w_1$) beams, $\mathcal{G}(\omega, \xi)$ is the Green's function of the sample in response to a point-pulsed source. In TDTR, the laser pulses are modulated by an electro-optical modulator (EOM). The time-domain modulated pulses can be represented as a frequency comb series in the frequency domain, such that the complex thermoreflectance response is calculated as a summation over the frequency-domain responses[2, 14]:

$$Z(\omega_0, \tau) = \sum_{n=-\infty}^{\infty} H(\omega_0 + n\omega_s) e^{in\omega_s \tau} \tag{2}$$

where $\omega_0$ is the modulation frequency of the pulsed pump laser (the base frequency), $\omega_s$ is the repetition rate, and $\tau$ is the probe delay time. The standard data reduction routine is to fit the measured phase signal $\phi = \arctan(\mathrm{Im}[Z\ or\ H]\ /\ Re[Z\ or\ H])$ or the ratio $R = -Re[Z\ or\ H]/Im\,[Z\ or\ H]$, because in this way the power-dependent pre-factor $\beta$ is normalized. Detailed formulations for data reduction can be seen in refs [2, 14, 63].

Both FDTR and TDTR can be extended to thermal property imaging using either the scanning approach or the wide-field approach [64]. The scanning approach builds the property image through repeating local measurements pixel by pixel. It can be most conveniently implemented by mounting the sample onto a mechianical or piezo translation stage[36, 38] (Fig. 1a) without changing the optics. Alternatively, scanning can be achieved by keeping the sample fixed but moving the laser spots using galvanometric mirrors[65]. A pair of galvanometer is used to control the tilting of scanning mirrors, such that the laser spot can be translated in the x- or y-directions (Fig. 1b). A "4*f*-telescope", consisting of two identical lenses separated by the sum of their focal lengths, is often implemented between the scanning mirrors and the objective to make sure the incident beams always pass through the center of the objective to minimize the beam distortion. On the other hand, the wide-field approach, usually referred to as transient thermoreflectance imaging (TTI), uses a laser light emitting diode (LED) as the probe source that can illuminate a large area, and a CMOS/CCD camera is used to capture thermoreflectance change across the field of view [66-67] (Fig. 1c). Compared with the scanning approach, the wide-field approach has the advantage of fast imaging speed, enabling high-resolution spatiotemporal observation of heat dissipation dynamics [68]. The wide-field approach also has the advantage of automatic alignment of optical and thermoreflectance images through the microscope, so that identifying the structural features and measuring their thermal image is straightforward. Nevertheless, the biggest challenge of the wide-field approach for thermal property mapping is complex data analysis. Instead of a simple Gaussian source that can be analytically modeled, extracting thermal properties using the wide-field approach requires numerical simulations to account for the heating pattern and structural features in the illuminated area [69]. Property inversion from wide-field imaging becomes more challenging for micro/nanostructured materials or devices, in which the thermal conductivity map $\kappa(x, y)$ are usually non-uniform. Therefore, the wide-field thermoreflectance is more suitable for direct temperature mapping rather than property mapping [70-71]. As shown in Fig. 1d, this challenge is recently mitigated by using structured illumination [72]: the pump laser is patterned using a digital micromirror device (DMD) so that thermal responses are constrained within smaller areas. Comparing the two imaging methods, the scanning approach has higher spatial resolution for property imaging (*i.e.,* the spot size) and is more often implemented for property mapping. However, the imaging speed of the scanning approach is limited, especially when performing high-resolution scans. As we shall discuss in later sections, the combination of AI with thermoreflectance imaging can not only overcome the limitations in measurement throughputs but also significantly improve the property mapping performance, such as overcoming the balance of imaging speed and accuracy, super-resolution imaging, and nonintrusive subsurface analysis.

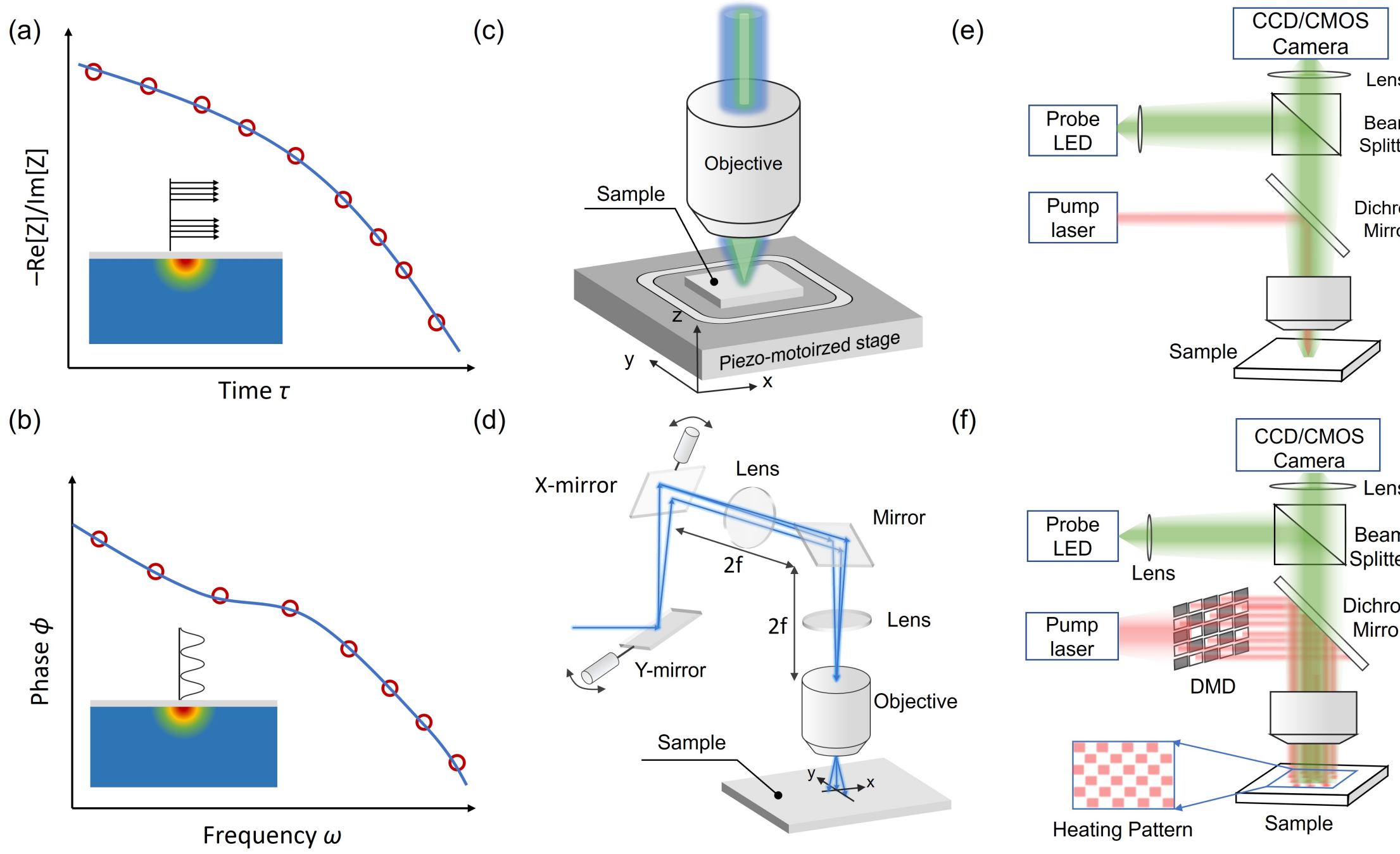

**Figure 1. Approaches of pump-probe thermoreflectance microscopy.** (a-b) Schematic of (a) TDTR and (b) FDTR detection and signals. (c-d) The scanning approach based on (c) piezo-motorized stages and (d) a galvanometric scanner with a 4*f*-telescope relay system. "4*f*" denotes that the distance between the scanning mirror and the objective is four times the focal length of two lenses. (e-f) The wide-field thermoreflectance imaging. (e) the pristine thermoreflectance imaging and (f) with structured illumination.

In the next few sections, we highlight the enhanced capability of mapping thermal properties by combining AI with pump-probe thermal reflectance imaging. First, ML models can serve as highly efficient surrogates for data reduction, which can significantly accelerate data reduction and uncertainty analysis. In addition to simple data reduction, ML techniques provide "smart sampling protocols", such that robust imaging can be generated with fewer measurements than required by the Shannon-Nyquist theorem. Finally, AI models enable unprecedented imaging performance, including super-resolution property mapping and nonintrusive subsurface analysis.

## 3. Accelerated Data Reduction

One of the most straightforward methods of combining AI with pump-probe imaging is to accelerate the data reduction by replacing analytical models with a more efficient AI surrogate model. Such AI-Surrogate-based data reduction can be particularly useful when the data fitting process needs to be repeated many times, such as property image extraction and Monte Carlo uncertainty analysis. Figure 2a shows the basic principles of accelerated data reduction using AI models. First, different combinations of experimental parameters are generated, including effective spot radii $w_{rms}$, modulation frequency $f$, thickness $d_{tr}$, specific heat $C_{tr}$ and thermal conductivity $\kappa_{tr}$ of metal transducers, interface conductance $G$, and the sample properties (thermal conductivity $\kappa$ and heat capacity $C$). These properties are used as input parameters to predict the pump-probe signal curves (typically the phase signal). Then an AI surrogate model can be trained based on these simulated properties and signal data under supervision, where the phase signal, the transducer properties, and experimental parameters (spot radii and modulation frequency) are labeled as input data, and the unknown properties (*e.g.* sample conductivity $\kappa$ and interface conductance $G$) are labeled as output data. One can often include some level of Gaussian noise in the predicted signal to increase the robustness of mapping between signals and the unknown properties [73]. After the AI model is trained, one can process the experimental signal pixel by pixel, and thermal property maps can be generated 2~3 orders of magnitude faster than direct data reduction using thermal models [73].

The highly efficient AI surrogate model also significantly accelerates the uncertainty propagation among multiple parameters using the Monte Carlo (MC) method. Since pump-probe data reduction relies on nonlinear least-squares regression, multi-parameter uncertainty analysis becomes rather complicated due to the coupling among different parameters [54]. MC can naturally incorporate multi-parameter coupling statistically. To implement the MC uncertainty analysis, each input parameter in the thermal model (except the modulation frequency *f*) is assumed to obey a normal distribution around its nominal value with a certain uncertainty level. The same set of experimental signals is then repeatedly processed by the AI model with the input parameters randomly selected, yielding new distributions of the output parameters as shown in Figure 2a [74]. Compared with analytical error propagation based on the Jacobi matrix [54], the MC uncertainty analysis does not require pre-determined best-fit values, and the effects of signal-to-noise ratio and fitting residuals are fully taken into account. Replacing the analytical thermal model with an AI surrogate model can significantly improve the computational speed. For example, without any parallelization, using the AI surrogate to perform 5000 MC fittings only takes ~ 12 seconds on a personal computer, while it can take a few hours using the analytical model due to the iterations in the regression and summation operators when computing the response functions.

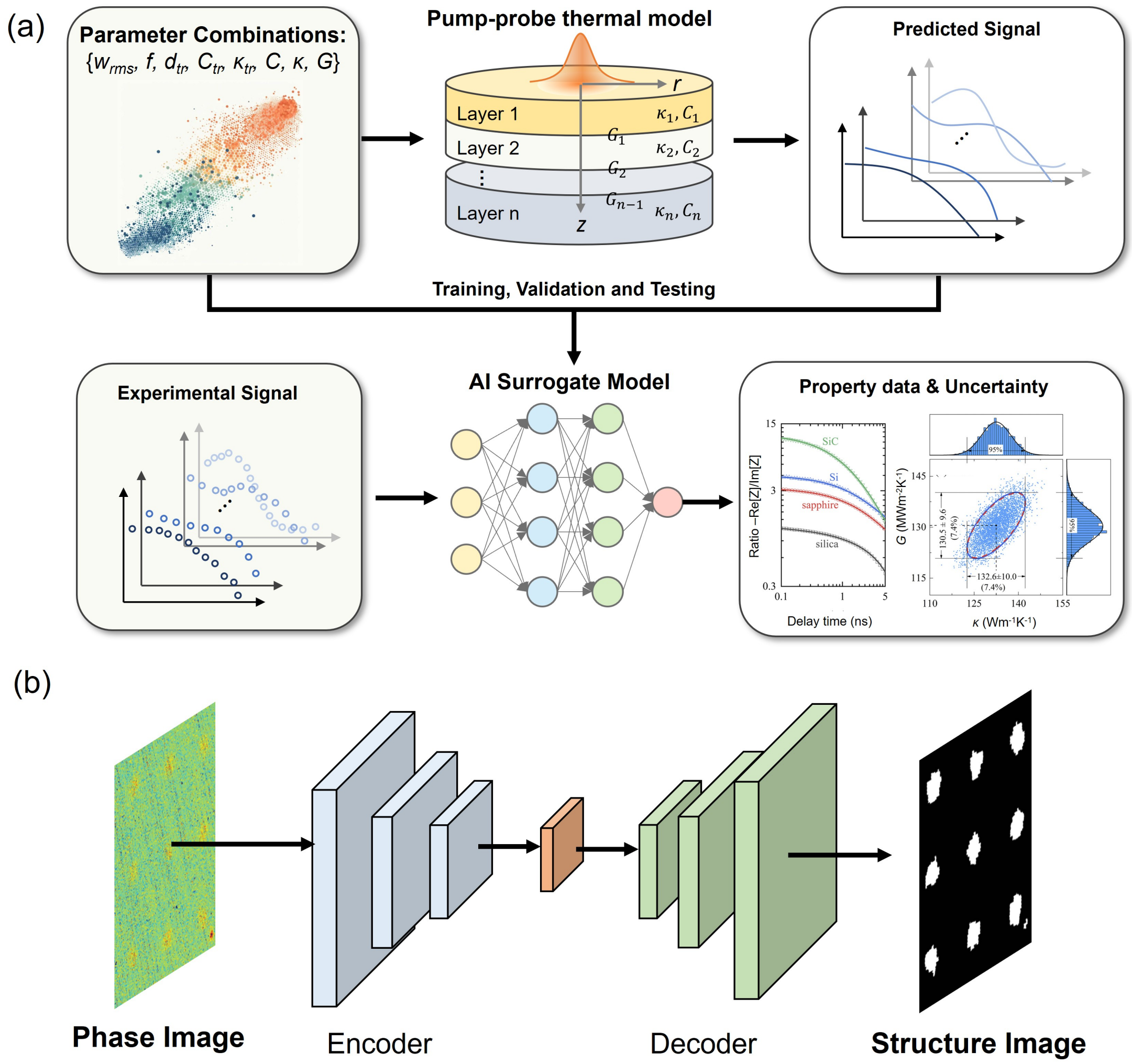

**Figure 2. High-throughput Surrogate Data Reduction.** (a) Property extraction using an AI surrogate model. Property data and the uncertainty figure are reproduced from Ref. [73]. (b) Full-field imaging processing. The phase image and structural image represent an array of indium bump bonds, reproduced from Ref. [75].

Figure 2b shows a further step of "full-field" image processing, where the experimental phase is treated as a whole image rather than individualized pixels to be processed repetitively. Convolutional neural networks (CNNs) containing encoders and decoders are particularly suitable for image processing with a large number of pixels [76]. The encoder blocks can be viewed as iterative mathematical transformations that automatically extract the key features from the phase image to the latent representation space with reduced dimensionality. The decoder blocks, on the other hand, are trained to map the key features back to property or structural distributions in the real space. A recent effort towards full-field image processing is demonstrated by Jarzembski et al. [75], where a CNN model is trained to identify the areas with indium bump bonds based on the thermal phase images. Such full-field qualitative analysis can substantially reduce the amount of data reduction cost while offering an overall evaluation of the property distribution. If one brutally performs quantitative data reduction, there are typically millions of pixels to be processed individually, which is very computationally cumbersome. Using the full-field qualitative approach, one can directly gain intuitions about spatial distributions of thermal properties, and then select representative locations to perform quantitative measurements. A natural future step is to directly reconstruct property images from the phase image in a full-field manner. To achieve this, structural-to-property image datasets need to be well-prepared, and embedding the physical laws of heat transfer into the AI image-processing algorithms could be pivotal to ensure the imaging accuracy.

## 4. Compressive sensing and image reconstruction

In addition to accelerated data reduction, combining machine learning or AI techniques with pump-probe imaging can even guide the "smart" experimental sampling. Yang *et al.* [77] recently demonstrated high-throughput property imaging using compressive sensing frequency-domain thermoreflectance (CS-FDTR) microscopy. Compressive sensing leverages the sparse feature of physical images in some representation space to enhance the throughput of measurements, such that the primary features embedded in the signal can be captured with fewer experiments than the Shannon-Nyquist sampling theorem [78]. For example, a realistic thermal property distribution represented in Fourier space is almost certain to exhibit a clustered distribution of large coefficients in low-frequency ranges. This is because there must be a minimal length scale, below which the thermal property variations are negligible, and the high-frequency components correspond to noise. For example, the minimum length for property variation can be viewed as the particle size in a nanocomposite. As a result, the number of nonzero components in the Fourier space is typically less than the total number of pixels, namely, a physically real image is generally sparse. According to the compressive sensing theory, the full image can be reconstructed robustly from the down-sampled raw image by solving a minimization problem [79].

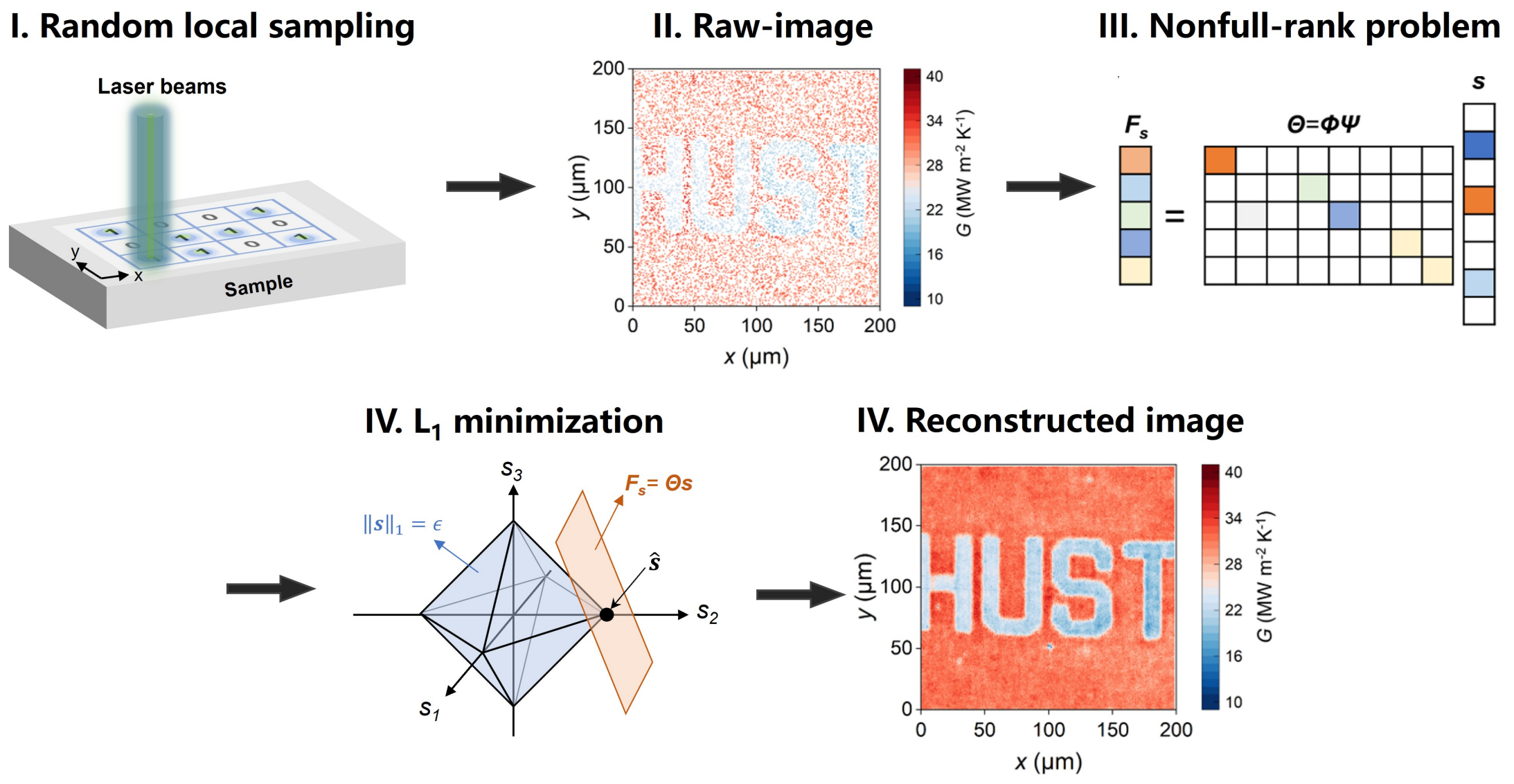

**Figure 3. Workflow of CS-FDTR property mapping.** Re-adapted from Ref. [77].

The workflow of CS-FDTR is outlined in Figure 3. First, random local measurements are taken to sample the field of view. Note that random sampling is preferred over uniform sampling, as the distance between randomly sampled locations can vary, which ensures both high-frequency and low-frequency features are captured. By fitting the phase data, a down-sampled raw image of thermal properties is generated with lots of "blind spots" represented by the white pixels. To fill in the blind spots with reasonable property values, one needs to solve an underdetermined linear problem $\boldsymbol{F}_s = \boldsymbol{\Theta} \cdot \boldsymbol{s}$, where $\boldsymbol{F}_s$ is the down-sampled raw property image, $\boldsymbol{\Theta}$ is the compressive sensing matrix embedding the random sampling locations and the Fourier transformation, $\boldsymbol{s}$ is the Fourier space representation of the image to be determined. The sparsest Fourier-space representation $\boldsymbol{s}$ can be evaluated by minimizing the LASSO cost function $\mathcal{L} = \|\boldsymbol{\Theta} \cdot \boldsymbol{s} - \boldsymbol{F}_s\|_2^2 + \lambda\|\boldsymbol{s}\|_1$, with $\lambda$ the regularization parameter [80]. The $L_1$-regularization term $\|\boldsymbol{s}\|_1$ ensures that the solution $\boldsymbol{s}$ can be found along some axis in the hyperdimensional Fourier-space, such that many components along many dimensions are zero. After obtaining the Fourier-space representation $\boldsymbol{s}$, the image can be reconstructed using a simple inverse Fourier transform. Using the CS-FDTR approach, thermal property mapping can be obtained by randomly measuring only 15%~50% of pixels, such that both the acquisition time and data processing can be reduced by a factor of 2~6 while maintaining the imaging accuracy. CS-FDTR demonstrated that the mutual constraint between imaging throughput and data quality can be broken by combining AI/ML with pump-probe imaging techniques.

## 5. Super-resolution thermal property mapping

The recent surge of artificial intelligence in computer vision has fundamentally changed how we approach image enhancement. In particular, single-image super-resolution (SISR) has moved from classical interpolation and deconvolution methods to deep learning architectures that can learn the complex mapping between low-resolution and high-resolution images. By training on large datasets of natural images, these models capture hierarchical features—edges, textures, and object structures—allowing them to reconstruct fine details that can be challenging to identify with limited resolution. This success has sparked great interest in applying similar techniques beyond photography, and recently to spatially resolved thermal property mapping.

A proof-of-concept super-resolution enhancement of sub-millimeter-scale thermal property mapping is demonstrated by Chen et al. [81], as shown in Figure 4, using a residual network (ResNet) model. The ResNet has become a cornerstone for modern computer vision [82] because it solves a longstanding problem in deep learning: the gradient of training parameters in the neural network vanishes as the number of hidden layers increases, making deep learning models difficult to train [83]. The ResNet has two most important components. First, element-wise sum operations are implemented. These operations add the input values to the output values, such that instead of directly learning to fit the pixel values, the ResNet learns how to minimize the “residual” part. This is especially important for thermal property mapping as it preserves the local property values of the input low-resolution image across many layers. Such element-wise sum operations also mitigate the problem of vanishing gradients. Second, the ResNet has skip connections, allowing the output of a layer to bypass a few layers and add to the neurons’ output. Such structural change allows gradients to directly flow through the network and mitigate the problem of vanishing gradients, enabling stable training even with a large number of layers. ResNet also incorporates some special layer structures to ensure stable super-resolution image reconstruction, such as convolutional layers[76] for extracting key features from the thermal property image, pixel shuffle for up-sampling the pixels[84], batch normalization to eliminated accumulation of drifts in parameters across the network[85], and parametric rectified linear units (PReLU) to ensure numerical activeness of neurons[86]. With such AI image enhancers, the resolution of property mapping has been improved by a factor of four, which can effectively reduce the amount of measurement time and data processing costs in large-area scans [81].

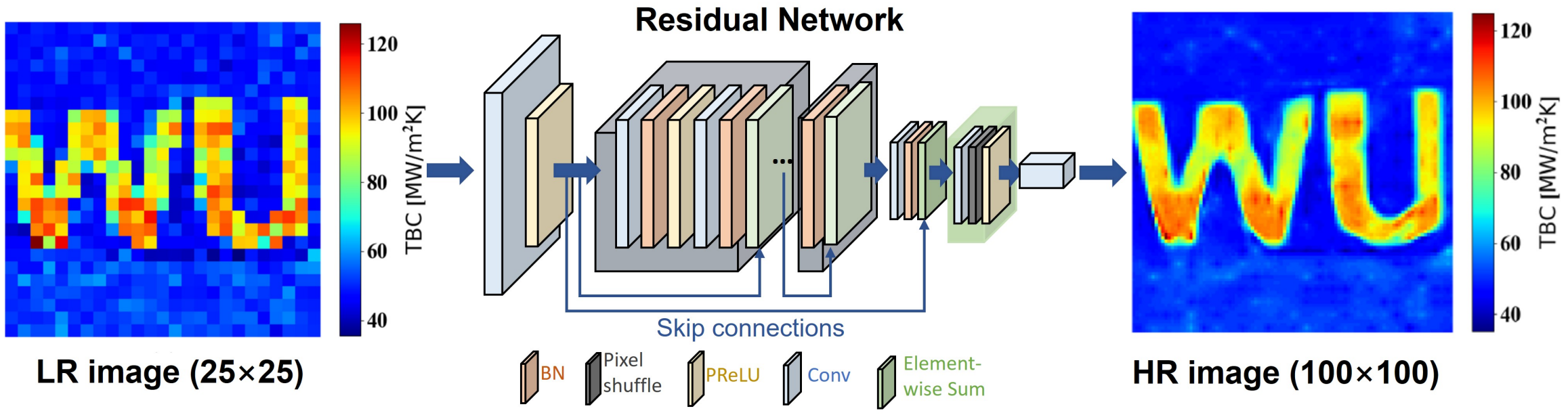

**Figure 4. AI-enabled super-resolution image reconstruction.** The low-resolution (LR) and reconstructed high-resolution (HR) images of the "WU"-patterned thermal conductance are reproduced from Ref. [81].

However, there are remaining physical limitations to the achievable imaging resolution at micro/nanoscales. With the piezo-motorized stages, the scanning precision can reach nanometer scales, which is even smaller than phonon mean free paths for lots of semiconductors[87-88] and is not the limitation for the spatial resolution of thermal properties. The physical limitation lies in the finite size of the laser beam and thermal diffusion, such that the experimentally measured map $k_{EXP}(x, y)$ is not exactly local. A schematic illustration is shown in Figure 5 using the interface between a high thermal conductivity particle embedded in a low-thermal conductivity matrix as an example. When performing pump-probe property imaging near the interfacial region, the Gaussian profile of the laser beam will cover both sides of the interface. In addition, there will be inevitable thermal "cross-talk": the temperature rise on one side is dependent on the thermal properties of the other side. When sensing thermal responses with a probe beam, the obtained response is an integration over the local temperature rise on both sides of the interface. Therefore, the measured property image will be "smeared". This physical smearing effect needs to be properly considered in order to push the spatial resolution beyond the beam size. While current AI models can generate super-resolution images, the physical smearing effect due to finite-sized pump-probe beams and the thermal cross-talk has not been fully incorporated into current AI-based algorithms for super-resolution imaging. Physics-informed AI techniques can be a fruitful direction to overcome the physical limitation on resolution, where the constitutive laws and governing partial differential equations are directly embedded into the loss function when training deep neural networks [89]. A recent numerical investigation by Odufisan et al. [90] treats the finite-sized heating and thermal cross-talk using a simplified Gaussian smearing, and showed that FDTR images can be enhanced if the microstructure can be fed to the AI model.

While the combination of AI with pump-probe imaging showed great potential in enhancing the mapping resolution, the following aspects remain to be addressed to truly surpass the physical resolution limit. First, how to properly quantify the spatial resolution of thermal property imaging should be investigated systematically, on par with the definition of the diffraction limit in optics. While the beam size is limited by the diffraction limit, the heat diffusion effects will further complicate the problem and blur the pump-probe thermal image. Second, embedding physical information with deep learning techniques specifically developed for image processing (such as CNNs and ResNets) can be investigated in the near future, especially for cases when the structural image is absent. Third, variations in thermal properties should be fully incorporated. To correctly model thermal responses, the diffusion term in the heat equation should be calculated as $\nabla \cdot (\kappa \nabla T) = \nabla \kappa \cdot \nabla T + \kappa \nabla^2 T$, but the current literature has neglected the gradients in thermal properties. Super-resolution thermal imaging, when fully realized, can serve as a novel technique for visualizing the structure-property relation for materials, interfaces, and devices not only at static conditions, but also in active systems during operations [91-92].

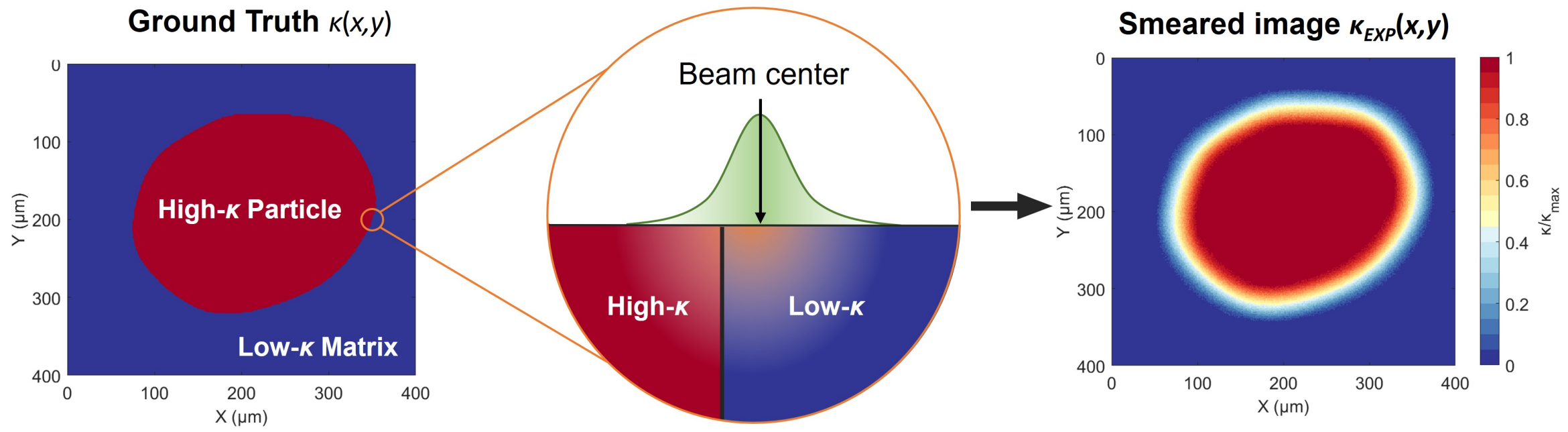

**Figure 5. Physical smearing effect due to finite beam size and heat diffusion.**

## 6. Non-intrusive subsurface analysis.

In addition to in-plane mapping of thermal properties, pump-probe thermal property microscopy has great potential for non-intrusive subsurface analysis. As shown in Figure 6a, under periodic laser heating, the excited temperature oscillations mainly occur within the penetration depth $L_p = \sqrt{\alpha/\pi f}$ [2], where $\alpha$ is the thermal diffusivity and $f$ the modulation frequency of the pump laser. When varying $f$, the measured thermal signals essentially reflects the heat diffusion at different depths. If thermal properties vary in the depth direction, the detected phase will deviate from the case with uniform property distribution. Therefore, the depth dependence is naturally embedded in the frequency-dependent thermal phase signals. For example, several works [57, 59, 93] leverage this feature to extract the depth dependence, which typically involves assuming or guessing a function $\kappa(z)$ with fitting parameters. However, such pre-knowledge, in most general cases, is not available before performing the measurements. The key challenge, therefore, lies in developing a data processing technique general enough for an arbitrary shape of $\kappa(z)$. AI or ML techniques are particularly suitable for dealing with such complex regression tasks, because of the ability to "learn" from training dataset and infer the shape from the experimental signal. This idea was first investigated by Xiang *et al.* [94] by implementing a simple kernel learning algorithm, and then Ikeda *et al.* [95] extended it to deep learning techniques for better scalability and accuracy.

The general workflow for AI-enabled depth profiling of thermal properties is shown in Figure 6b. First, a training database is constructed for the ML model to learn about the mapping from $\kappa(z)$ to the thermal phase signal $\phi(f)$. To do this, various shapes of $\kappa(z)$ are first assumed, along with different combinations of experimental parameters, including the effective beam radius $w_{rms}$, transducers' thicknesses ( $d_{tr}$ ) and thermal properties($\kappa_{tr}, C_{tr}$), interfacial conductance $G$, and heat capacities. These data are fed to the pump-probe model to calculate the corresponding phase signals. To construct the database, experimental parameters and the phase signals are labeled as input data $\boldsymbol{X}$, and the unknown properties $\kappa(z), G$ are labeled as output data $\boldsymbol{Y}$ for the ML model. The database is separated into three subsets: the training set to parameterize the ML model, the validation set for monitoring the performance evolution during the training, and the testing set for evaluating the accuracy of reconstructing $\kappa(z)$ of the trained model. After the ML model is trained, the experimental phase signal $\phi_{exp}$ is fed as input for the ML model, and the model will infer the most probable $\kappa(z)$. Fig. 6c-d shows two typical $\kappa(z)$ shapes from the testing set, which were not seen by the ML model in the training and validation process, and the excellent agreement between these sets demonstrated that ML can extract $\kappa(z)$ accurately even when it cannot be described by simple analytical functions. Fig. 6e shows the performance of depth-profiling of thermal conductivity for an ion-irradiated silicon sample. The reconstructed signal shows suppressed thermal conductivity for depths from 0.1 µm to ~0.8 µm, which agrees well with the density peak of irradiated ions. Note that no information about the ion density distribution is provided to the ML model. This opens up the possibility of non-intrusive depth-profiling of thermal properties, which can be used to detect defects or buried structures below the sample surface.

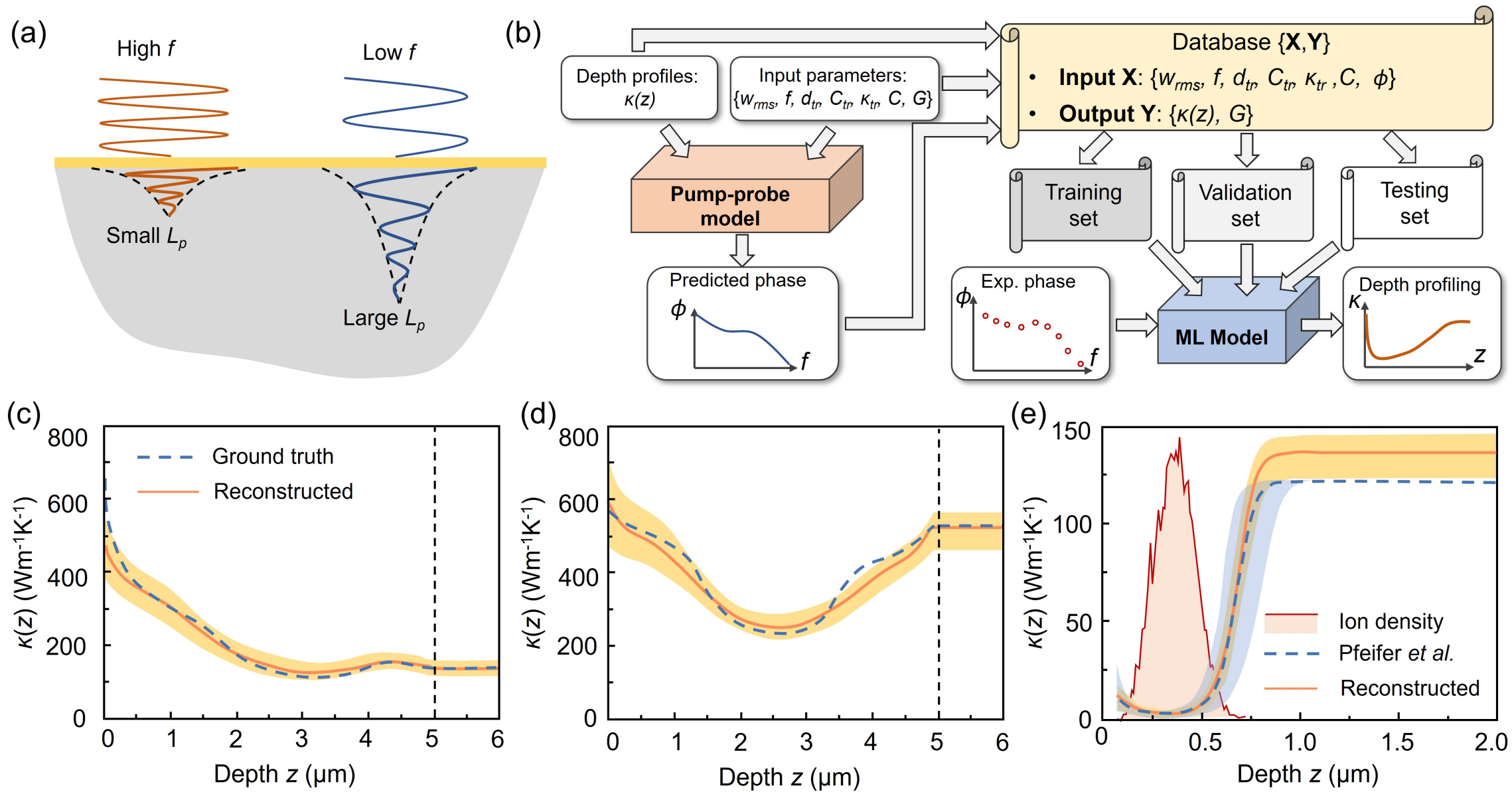

**Figure 6. Nonintrusive depth profiling enabled by machine learning.** (a) Penetration depth $L_p$ modulated by heating frequency $f$. (b) Workflow of depth profiling of thermal conductivity $\kappa(z)$ using machine learning models. A pump-probe thermal model is used to generate phase signals and construct a database for training the machine learning (ML) model. (c-d) Reconstructed $\kappa(z)$ compared with the ground truths in the testing set. (e) Machine learning reconstructed $\kappa(z)$ of ion-irradiated silicon; the reference $\kappa(z)$ and ion density (peak value normalized to unity) measured by Pfeifer *et al.* are taken from ref.[59]. (c-e) is readapted from Ref. [94].

In addition to thermal conductivity profiling, analysis of buried heterogeneous interfaces is a significant aspect of ensuring optimal heat dissipation and mechanical stability of high-power integrated circuits. When the heat penetrates to the buried interface, the thermal phase signal will be sensitive to the interfacial thermal conductance (Fig. 7a-c). Therefore, pump-probe thermal imaging can serve as an effective tool for analyzing the interfacial quality of heterogeneous bonding. Although one can perform pixel-by-pixel data reduction to identify the bonded, partially bonded, and unbonded regions, ML techniques can automatically learn from the experimental signals and uncover the spontaneous clustering of experimental data, even when the model is unaware of the bonding qualities. An example of data-clustering analysis is shown in Fig. 7d, using the $k$-means method[50]. Such clustering analysis allows researchers to select representative data points for further detailed analysis. As shown in Fig. 7d, different data clusters exhibit very different thermal phase signal curves, which correspond to physical regions with different interfacial conductance. Furthermore, the spatial distribution of interfacial conductance can be rapidly reconstructed by weighted average over different clusters. To achieve this, one can decompose the phase signal as a linear combination of data clusters: $\phi_{k-means}(f,x,y) = \sum_{j=1}^{k} c_j(x,y)\phi_j(f)$, where $k$ is the total number of clusters, and $\phi_j$ is the representative signal measured in the $j$-th cluster. The weights $c_j(x,y)$ can be obtained by fitting the $\phi_{k-means}$ to the experimental image $\phi_{exp}(f,x,y)$ using least-squares regression. Then, the interfacial conductance mapping can be generated as $G_2(x,y) = \sum_{j=1}^{k} c_j(x,y)G_j$, where $G_j$ denotes the interfacial property obtained by fitting the signal of the $j$-th cluster, as shown in Fig. 7e. The reconstructed mapping is shown in Fig. 7f. Such clustering analysis can automatically identify the data grouping pattern without human intervention and provide physical insights about the bonding quality distribution of the buried interface with a nearly two-order-of-magnitude speed improvement over pixel-by-pixel data processing.

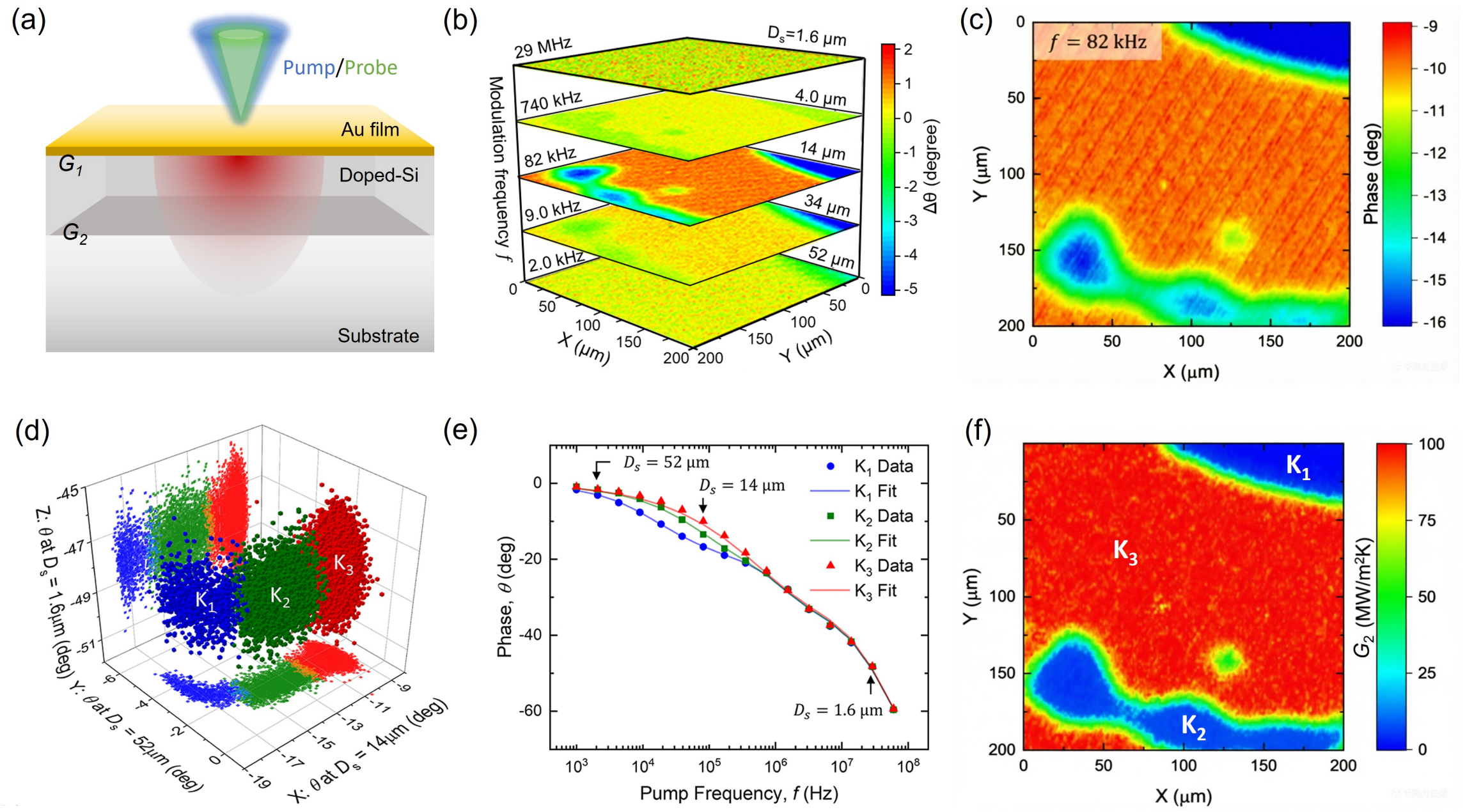

**Figure 7. Data clustering enabled rapid analysis of buried interface conductance.** (a) Geometry of pump-probe conductance imaging of the buried interface $G_2$. (b) Experimental phase image at different modulation frequencies. $D_s$ denotes the sensing depth. (c) Phase image at $f = 82$ kHz. (d) Data clustering of the phase images. (e) Phase signals and best fits corresponding to the data clusters ($K_1$, $K_2$, and $K_3$). (f) Interfacial conductance image. Major regions corresponding to the data clusters are labeled, with $K_1$ and $K_2$ corresponding to low-conductance regions (weakly bonded) and $K_3$ corresponding to the high-conductance and strongly bonded area. Reproduced from ref. [50].

## 7. Outlooks

The above discussion highlights the recent efforts and advancement of combining AI with pump-probe thermal property imaging to achieve unprecedented measurement throughput, accuracy, spatial resolution, and the capability of non-intrusive subsurface analysis. Here we offer outlooks for a few promising future directions.

(1) Current implementation of AI with pump-probe thermal imaging remains within the diffusive heat conduction regime. Phonon transport, however, might exhibit various transport dynamics, including coherence, localization, hydrodynamics, and quasi-ballistics[87]. Future work including rigorous microscopic phonon transport physics, such as the Boltzmann transport equation[96] or Wigner transport equation[97], into the AI models can extend the scope of AI-enhanced thermal property imaging to AI-driven phonon spectroscopy, enabling high-fidelity extraction of frequency-resolved phonon properties.

(2) With the development of in-plane mapping and depth-profiling methods, a natural yet challenging next step is to push toward three-dimensional thermal property imaging. There exist both experimental and numerical challenges. On the experimental side, in-plane and cross-plane imaging resolutions become convoluted, and determining the optimal experimental condition can become very difficult. For example, to achieve higher in-plane spatial resolution, tightly focused laser beams and high modulation frequencies are desirable, but these requirements will severely limit the penetration depth, leading to vanishing sensitivity to thermal responses in deeply buried regions [56]. On the numerical side, the issue of multiple-parameter coupling can also severely complicate the image reconstruction processes, especially for anisotropic materials and structures. Pushing towards three-dimensional thermal property imaging can also impose challenges in scalability and robustness when training AI models.

(3) The status quo of combining AI with pump-probe thermal imaging is primarily at the data-processing level, providing very few guidelines on how to perform measurements, with only a few exceptions [50, 77]. In addition to solving complex regression and image reconstruction tasks, AI can also robustly perform feature selection tasks. For thermal property imaging, AI feature selection algorithms can be used to analyze the irreducible, independent, and highly sensitive parameters determining the pump-probe signal. Further, with the substantial advancements in large language models (LLMs) [98-100] towards artificial general intelligence, LLMs might be further trained as "AI experts" that can learn from existing literature and trial experimental results, and help the human researchers design the proper experimental protocol when investigating a specific sample [101].

## Acknowledgment

This work is supported by the National Key R&D Program (2022YFA1203100), and the National Natural Science Foundation of China (NSFC, Grant Nos. T2588301 and U25D9006). The authors declare no conflict of interest.